\documentclass[aps, prd
, twocolumn
, nofootinbib 
, longbibliography
]{revtex4-1}
\usepackage{graphicx}
\usepackage[caption=false]{subfig}
\usepackage{mathrsfs}
\usepackage{amsmath,amssymb}
\usepackage{bm}
\usepackage{braket}
\usepackage{listings}
\usepackage{cases}
\usepackage{comment}
\usepackage{soul}
\usepackage{cancel}
\usepackage{cases}
\usepackage[utf8]{inputenc}
\usepackage{url}
\usepackage{longtable}
\usepackage[normalem]{ulem}
\usepackage{xspace}

\usepackage[colorlinks=true
,urlcolor=blue
,anchorcolor=blue
,citecolor=blue
,filecolor=blue
,linkcolor=blue
,menucolor=blue
,linktocpage=true
,pdfproducer=medialab
,pdfa=true
]{hyperref}

\newcommand{\dif}[2]{\frac{\mathrm{d} #1}{\mathrm{d} #2}}
\newcommand{\pdif}[2]{\frac{\partial #1}{\partial #2}}

\newcommand{\dd}{\mathrm{d}}

\newcommand{\ee}{\mathrm{e}}

\newcommand{\Mpl}{M_\mathrm{Pl}}
\newcommand{\lpl}{l_\mathrm{Pl}}
\newcommand{\ns}{n_{{}_\mathrm{S}}}

\newcommand{\eff}{\mathrm{eff}}

\newcommand{\infl}{\mathrm{inf}}

\newcommand{\hill}{\mathrm{hill}}
\newcommand{\fNL}{f_{\mathrm{NL}}}

\newcommand{\uf}{\mathrm{f}}

\newcommand{\ui}{\mathrm{i}}

\newcommand{\calO}{\mathcal{O}}

\newcommand{\calP}{\mathcal{P}}

\newcommand{\bae}[1]{\begin{align} #1 \end{align}}
\newcommand{\bce}[1]{\begin{cases} #1 \end{cases}}
\newcommand{\dps}{\displaystyle}
\newcommand{\bfe}[4]{
\begin{figure} 
	\centering
	\includegraphics[#1]{#2}
	\caption{#3}
	\label{#4}
\end{figure}}

\begin{document}
\title{Escape from the swamp with spectator}
\date{\today}

\author{Kazuhiro Kogai}
\author{Yuichiro Tada}
\email{tada.yuichiro@e.mbox.nagoya-u.ac.jp}
\affiliation{Department of Physics, Nagoya University, Nagoya 464-8602, Japan}

\begin{abstract}
In the context of string theory, several conjectural conditions have been proposed for low energy effective field theories not to be in \emph{swampland}, the UV-incomplete class.
The recent ones represented by the de Sitter and trans-Planckian censorship conjectures in particular seem to conflict with the inflation paradigm of the early universe.
We first point out that scenarios where inflation is repeated several times (multi-phase inflation) can be easily compatible with these conjectures.
In other words, we relax the constraint on the single inflation for the large scale perturbations to only continue at least around 10 e-folds.
In this context, we then investigate if a spectator field can be a source of the almost scale-invariant primordial perturbations on the large scale.
As a consequence of such an isocurvature contribution, the resultant perturbations exhibit the non-vanishing non-Gaussianity in general. 
Also the perturbation amplitude on smaller scales can be completely different from that on the large scale due to the multiplicity of inflationary phases.
These signatures will be a smoking gun of this scenario by the future observations.
\end{abstract}

\maketitle

\section{Introduction}

The inflation paradigm has so far achieved great success as the scenario of the early universe.
It naturally realizes the globally homogeneous universe, and moreover can be a source of local cosmic structures as confirmed by observations of e.g. the cosmic microwave background (CMB)~\cite{Akrami:2018odb} and the Lyman-alpha forest~\cite{Bird:2010mp}.
Though the existence of the inflationary phase itself is strongly supported, its concrete mechanism is however still unclear because of the lack of information about characteristic features such as the primordial tensor perturbations and the non-Gaussianity of scalar perturbations.
Some novel approaches might be required not only observationally but also theoretically.

In the context of string theory, the concept of \emph{landscape} and \emph{swampland} has been attracting attentions on the other hand.
While string theory is thought to be able to realize vast classes of low energy effective theories (landscape)~\cite{Douglas:2003um},
it was suggested that some effective field theories (EFTs) might be incompatible with the UV completion even though they look consistent at the low energy (swampland)~\cite{Vafa:2005ui} (see also Ref.~\cite{Palti:2019pca} for a review).
Several conjectural conditions, e.g., the weak gravity conjectures~\cite{ArkaniHamed:2006dz} and the distance conjectures~\cite{Ooguri:2006in}
have been proposed for landscape EFTs to satisfy, and considered as attractive suggestions to low energy physics from high energy string theory.
In particular, the recent de Sitter (dS) conjecture~\cite{Obied:2018sgi,Garg:2018reu,Ooguri:2018wrx} and trans-Planckian censorship conjecture (TCC)~\cite{Bedroya:2019snp,Bedroya:2019tba} tightly constrain the scenario of inflation.
Leaving their details aside for now, one can briefly say that they tend to disfavor the long-lasting inflationary universe, while a sufficient expansion ($\sim50\text{--}60$ e-folds) is required for a successful cosmology.
Taking it seriously, many authors have investigated possible loop holes.
For example, 
multi-field models~\cite{Achucarro:2018vey,Damian:2018tlf,Achucarro:2019pux,Bjorkmo:2019aev,Fumagalli:2019noh,Lynker:2019joa,Bjorkmo:2019fls,Aragam:2019omo,Bravo:2019xdo,Chakraborty:2019dfh}, 
excited initial state~\cite{Brahma:2018hrd,Ashoorioon:2018sqb}, 
warm inflation~\cite{Das:2018hqy,Motaharfar:2018zyb,Das:2018rpg,Kamali:2018ylz,Bastero-Gil:2018yen,Kamali:2019ppi,Dimopoulos:2019gpz,Bastero-Gil:2019gao,Das:2019hto,Kamali:2019xnt,Goswami:2019ehb,Rasheed:2020syk}, 
brane inflation~\cite{Lin:2018kjm,Lin:2018rnx,Sabir:2019bsh,Lin:2019fdk}, 
gauge inflation~\cite{Park:2018fuj}, 
non-minimal coupling to gravity~\cite{Yi:2018dhl,Cheong:2018udx}, 
modified gravity~\cite{Artymowski:2019vfy}, quantum correction~\cite{Holman:2018inr,Geng:2019phi}
etc. are discussed in the light of the dS conjecture (see also the references in Ref.~\cite{Mizuno:2019pcm}).
TCC in inflationary models are discussed in e.g. Refs.~\cite{Tenkanen:2019wsd,Brahma:2019unn,Schmitz:2019uti,Kadota:2019dol,Brahma:2019vpl,Lin:2019pmj,Saito:2019tkc,Brahma:2020cpy}.

Another simple solution is repeating inflationary phases many times which we dub multi-phase inflation. 
Though each phase cannot continue long, the required expansion can be reached in total with a sufficient number of inflation.
String theory generally provides ubiquitous scalar fields, which also supports the scenario that multiple scalar fields realize multiple phases of inflation.

Multi-phase inflation however has a drawback in perturbations.
To see this clearly, let us assume that each inflation phase is governed by (effectively) single field for simplicity. In this case, the dS conjecture claims that either of the absolute values of the two slow-roll parameters cannot be small as
\bae{\label{eq: dSC in single field}
    \epsilon_V\!=\!\frac{\Mpl^2}{2}\left(\frac{V^\prime}{V}\right)^2\!\gtrsim\calO(1), \quad \text{or} \quad \eta_V\!=\!\Mpl^2\frac{V^{\prime\prime}}{V}\!\lesssim-\calO(1),
}
with any possible field value.
$\Mpl=\sqrt{1/8\pi G}$ is the reduced Planck mass and $V$ is a scalar potential for a canonically normalized inflaton.
It does not necessarily prohibit inflation as long as $\epsilon_V\ll1$.
However the spectral index of primordial curvature perturbations, 
which is roughly estimated as
\bae{
	\ns-1\approx-6\epsilon_V+2\eta_V,
} 
by naively adopting the slow-roll approximation,\footnote{We use a symbol~$\approx$ when we formally use the slow-roll approximation but the slow-roll parameters are not small enough.} 
is never small in this case unless an accidental cancellation.
The observations of the cosmic microwave background (CMB) have already revealed the primordial perturbations are almost scale-invariant as 
$\ns=0.965\pm0.004$~\cite{Aghanim:2018eyx}.
Thus the naive multiple single-field inflation scenario is in serious conflict with observations (see also e.g. Ref.~\cite{Kinney:2018nny}).

Relaxing the single-field assumption may solve this problem. 
For example, though the background dynamics of each inflation keeps assumed to be determined by single field for simplicity, some spectator fields can contribute to perturbations, like as the curvaton mechanism~\cite{Enqvist:2001zp,Lyth:2001nq,Moroi:2001ct} or the modulated reheating scenario~\cite{Dvali:2003em,Kofman:2003nx}.
In these cases, the expression of the spectral index is modified as
\bae{
    \ns-1\simeq-2\epsilon_H+\frac{2}{3}\frac{m_{\sigma,\eff}^2}{H^2},
}
where $\epsilon_H=-\dot{H}/H^2$ is the first slow-roll parameter and $m_{\sigma,\eff}$ represents the effective spectator mass during inflation.
Thus, as long as $\epsilon_H\ll1$, a slightly tachyonic spectator $m_{\sigma,\eff}^2\sim-0.05H^2$ could be compatible with the CMB observation, 
even the inflaton satisfying the condition~(\ref{eq: dSC in single field}) by a large $|\eta_V|>1$.\footnote{Note that the dS conjecture requires at least one unstable direction for any $V>0$ point (see Eq.~(\ref{eq: dS conjecture}) for the original statement). 
Thus as long as the inflaton satisfies the condition~(\ref{eq: dSC in single field}), adding stable spectators does not matter.}
It is also noted that the extra degrees of freedom (d.o.f.) during inflation generally leaves non-vanishing non-Gaussianity in perturbations, which can be a testability of this scenario.

In this paper, we investigate such a spectator scenario, allowing that the CMB scale inflation does not continue enough for our whole observable universe in the light of multi-phase inflation and swampland conjectures. 
In Sec.~\ref{sec: multi-phase inflation and swampland conjecture}, the compatibility of the multi-inflation scenario with swampland conjectures is discussed. 
Numerically calculated spectator perturbations in a specific example are shown in Sec.~\ref{sec: curvaton scenario in multi-phase inflation}.
In Sec.~\ref{sec: conclusions}, we discuss whether the curvaton or modulated reheating scenario can consistently convert the spectator perturbations into the adiabatic curvature perturbations. Observational crosschecks of our scenario are also mentioned.
We adopt the natural unit $\hbar=c=1$ throughout this paper.

\section{Multi-phase inflation and swampland conjecture}\label{sec: multi-phase inflation and swampland conjecture}

String theory has a generic view of \emph{landscape}~\cite{Douglas:2003um}, that is, various types of low energy EFT can be given in a stringy (UV complete) framework. However it has been also suggested that some EFTs may be in \emph{swampland}~\cite{Vafa:2005ui}, i.e., they seem to have no problem at low energy but are not actually UV complete.
Several conditions have been so far proposed for EFT not to be in swampland. Inflationary models, which are often described in a form of EFT, are not an exception to be constrained by such conditions. 
For example, the distance conjecture~\cite{Ooguri:2006in} suggests that the canonical excursion of any scalar fields during inflation cannot exceed order unity in the Planck unit:
\bae{\label{eq: distance conjecture}
    \Delta\phi\lesssim\Mpl.
}
The dS conjecture~\cite{Obied:2018sgi} (and its refined version~\cite{Garg:2018reu,Ooguri:2018wrx}) prohibits a flat plateau in a scalar potential $V$, requiring the condition
\bae{\label{eq: dS conjecture}
    |\nabla V|\Mpl\geq cV, \quad \text{or} \quad \min(\nabla_I\nabla_JV)\Mpl^2\leq -c^\prime V,
}
at any field-space point for some universal constants $c,c^\prime>0$ of order unity. Here $|\nabla V|=\sqrt{G^{IJ}V_IV_J}$ is the invariant norm of the gradient with the inverse metric $G^{IJ}$ of the target space for all scalar fields including spectators if exist.
$\min(\nabla_I\nabla_JV)$ is the minimum eigenvalue of the Hessian $\nabla_I\nabla_JV$.
This conjecture claims in other words that there exists at least one unstable direction for any $V>0$ point.
Particularly in the canonical (effective) single-field case, the inflaton should be unstable as
\bae{\label{eq: single-field dSC}
    \epsilon_V=\frac{\Mpl^2}{2}\left(\frac{V^\prime}{V}\right)^2\geq\frac{c^2}{2}, \quad \text{or} \quad \eta_V=\Mpl^2\frac{V^{\prime\prime}}{V}\leq-c^\prime.
}
Though it forbids the slow-roll inflation, the accelerated expansion of the universe ($\epsilon_H=-\dot{H}/H^2<1$) itself is not necessarily prohibited as long as $\epsilon_V\ll1$.
However, even in such a case, the large negative value of $\eta_V$ implies an exponential grow of $\epsilon_H$ as
\bae{\label{eq: SR evolution of eH}
    \frac{\dd}{H\dd t}\log\epsilon_H\approx-2\eta_V,
}
and therefore inflation cannot continue so long.
Finally TCC~\cite{Bedroya:2019snp,Bedroya:2019tba} claims that the sub-Planckian perturbation will never cross the horizon by expansion, that is,
\bae{\label{eq: TCC}
    \frac{a(t)}{a_\mathrm{ini}}\lpl<\frac{1}{H(t)},
}
at any time $t$ with an initial scale factor $a_\mathrm{ini}$. $\lpl=\sqrt{G}$ is the Planck length. 
It implies that the inflation duration is strongly suppressed, depending on the inflation energy scale.

As the dS conjecture at least allows short inflation, one sees that repeated short inflation can give enough expansion for our observable universe
consistently with the dS conjecture~(\ref{eq: dS conjecture}).
Such repetition of inflation can be dynamically realized e.g. by coupling many single-field hilltop-type potentials with the Planck-suppressed operators~\cite{Kumekawa:1994gx,Izawa:1997df,Kawasaki:1997ju,Tada:2019amh}:
\bae{
    V_{\inf}=\sum_iV_{\hill,i}(\phi_i)+\sum_{i\neq j}\frac{1}{2}c_{ij}V_{\hill,i}(\phi_i)\frac{\phi_j^2}{\Mpl^2},
}
though the later discussion does not depend on the repetition mechanism so much.
The subscripts $i,j,\cdots$ label the phases of inflation and the corresponding inflaton fields. 
Each energy scale is assumed to be well hierarchical as 
$V_{\hill,i}(0)\gg V_{\hill,i+1}(0)$ for simplicity.
Also the positive coupling constants $c_{ij}$ are naturally supposed to be order unity.\footnote{If it is negative, the corresponding scalar is not stabilized and cannot play a role of inflaton, so that it is safely excluded.}
In this setup, each inflaton field is stabilized to its potential top at first through these couplings.
During the phase-$i$, the potential $V_{\hill,j}$ for $j<i$ is well decayed out and thus $\phi_{i+1}$ is stabilized only by $V_{\hill,i}$ because any other potential is negligible due to the scale hierarchy.
After the phase-$i$, the field $\phi_i$ oscillates and $V_{\hill,i}$ is diluted by the expansion of the universe. 
When $V_{\hill,i}$ gets as small as $V_{\hill,i+1}(0)$, the potential $V_{\hill,i}$ cannot stabilize $\phi_{i+1}$ any longer and then the phase-$(i+1)$ inflation is turned on.
In this way, inflation is automatically repeated. Each phase is driven by effectively single field.\footnote{One may avoid the exact maximum of the potential (symmetric point) for the stabilizing point so that the inflatons' dynamics is determined only by the background evolution, or otherwise the quantum diffusion significantly affects the dynamics. In this paper, we only treat the phase-0 explicitly and the initial value of $\phi_0$ is shifted by hand for simplicity.}

On top of each $V_{\hill,i}$ where the single-field hilltop inflation occurs, the second condition of the single-field dS conjecture~(\ref{eq: single-field dSC}) 
should be satisfied because the first condition is violated in order for an accelerated expansion ($\epsilon_H\sim\epsilon_V\ll1$).
The distance conjecture~(\ref{eq: distance conjecture}) is also satisfied in general under this assumption.
In other parts of $V_{\hill,i}$, the first condition can be satisfied.
Therefore, even in the full multi-inflaton target space, the dS conjecture~(\ref{eq: dS conjecture}) is satisfied along the trajectory realized in inflation.
The potential may be modified to satisfy the condition at other points but they are irrelevant to the inflationary dynamics.
TCC is also much relaxed in the multi-phase inflation scenario as we will see later (see also Refs.~\cite{Mizuno:2019bxy,Berera:2019zdd,Li:2019ipk}).
Thus the swampland conjectural conditions can be satisfied simply by assuming that inflation is repeated many times.
For convenience, let the phase-$0$ correspond with the CMB scale.
Only the phase-0 is constrained also by observations because it is responsible for the CMB scale.
Hereafter we merely assume that the phase-0 is governed by effectively single-field hilltop inflation and followed by repeated inflation ($i=1,2,3,\cdots$) 
without specifying the details of following inflation ($i>0$) and the existence of preinflation ($i<0$).
In the rest of this section, we discuss the required condition for the phase-0 under this assumption.

For a concrete discussion, let us first expand $V_{\hill,0}$ as
\bae{
    V_{\hill,0}(\phi_0)=\Lambda^4-\frac{1}{2}\kappa\Lambda^4\frac{\phi_0^2}{\Mpl^2}+\cdots.
}
Both the Hubble parameter and the second slow-roll parameter are almost constant as $H\simeq\Lambda^2/\sqrt{3}\Mpl$ and $\eta_V\simeq-\kappa$ during the phase-0.
The dS conjecture~(\ref{eq: single-field dSC}) requires $\kappa\gtrsim1$.
Once the time evolution of $H$ is neglected (i.e. $\epsilon_H\ll1$), the background equation of motion (EoM)
\bae{\label{eq: background EoM}
    0=\ddot{\phi}_0+3H\dot{\phi}_0+V_{\hill,0}^\prime\simeq\ddot{\phi}_0+3H\dot{\phi}_0-3\kappa H^2\phi_0,
}
has an analytic solution as
\bae{
    \phi_0(t)=\phi_0(t_0)\exp\left[\frac{1}{2}\left(-3+\sqrt{9+12\kappa}\right)H(t-t_0)\right].
}
$t_0$ is some initial time.
Noting that $\epsilon_V\simeq\kappa^2\phi_0^2/2\Mpl^2$, one finds the evolution equation
\bae{\label{eq: evolution of eV}
    \dif{}{N}\log\epsilon_V\simeq-3+\sqrt{9+12\kappa}.
}
with use of the e-foldings $\dd N=H\dd t$ as the time variable. This is the generalization of the slow-roll equation~(\ref{eq: SR evolution of eH}) for large $|\eta_V|>1$.
It shows an exponential grow of $\epsilon_V$ and therefore it should be significantly small at $t_0$ so that the phase-0 continues sufficiently.

On the other hand, $\epsilon_V$ has a lower limit depending on the inflation scale $\Lambda$.
That is because the curvature perturbations generated by $\phi_0$ should be smaller than the observed value 
$\calP_\zeta\simeq2\times10^{-9}~$\cite{Aghanim:2018eyx} to utilize the spectator scenario. 
The curvature perturbations given by the inflaton is estimated as
\bae{
    \calP_{\zeta_{\phi_0}}\approx\frac{1}{24\pi^2\Mpl^4}\frac{\Lambda^4}{\epsilon_V}\lesssim2\times10^{-9}.
}
It reads a lower bound on $\epsilon_V$ at the onset of the observable scale $k^{-1}\simeq14\,\mathrm{Gpc}$ as
\bae{
    \epsilon_V(t_{14\,\mathrm{Gpc}})\gtrsim\frac{1}{24\pi^2(2\times10^{-9})}\frac{\Lambda^4}{\Mpl^4}.
}
Combining this lower limit and the evolution equation~(\ref{eq: evolution of eV}), one finds an upper bound on $\Lambda$ depending on $\kappa$ for the phase-0 to continue enough.
In Fig.~\ref{fig: lambda kappa}, we show this bound, requiring 15 e-folds from the onset of the observable scale $t_{14\,\mathrm{Gpc}}$ to the end of phase-0 
as a conservative line. 
It is numerically checked by solving the full background EoM (the first equation of (\ref{eq: background EoM})).
One should here recall that there is also a general lower limit on $\Lambda$, that is, inflation should be completed well before the big-bang nucleosynthesis (BBN) era $\sim1\,\mathrm{MeV}$. Conservatively it reads $\Lambda\gtrsim1\,\mathrm{MeV}$, which is also shown in Fig.~\ref{fig: lambda kappa}.
Combining them, one finds that the potential curvature $\kappa$ cannot be larger than $\kappa\lesssim18$ in the phase-0 (the CMB scale).

\bfe{width=0.9\hsize}{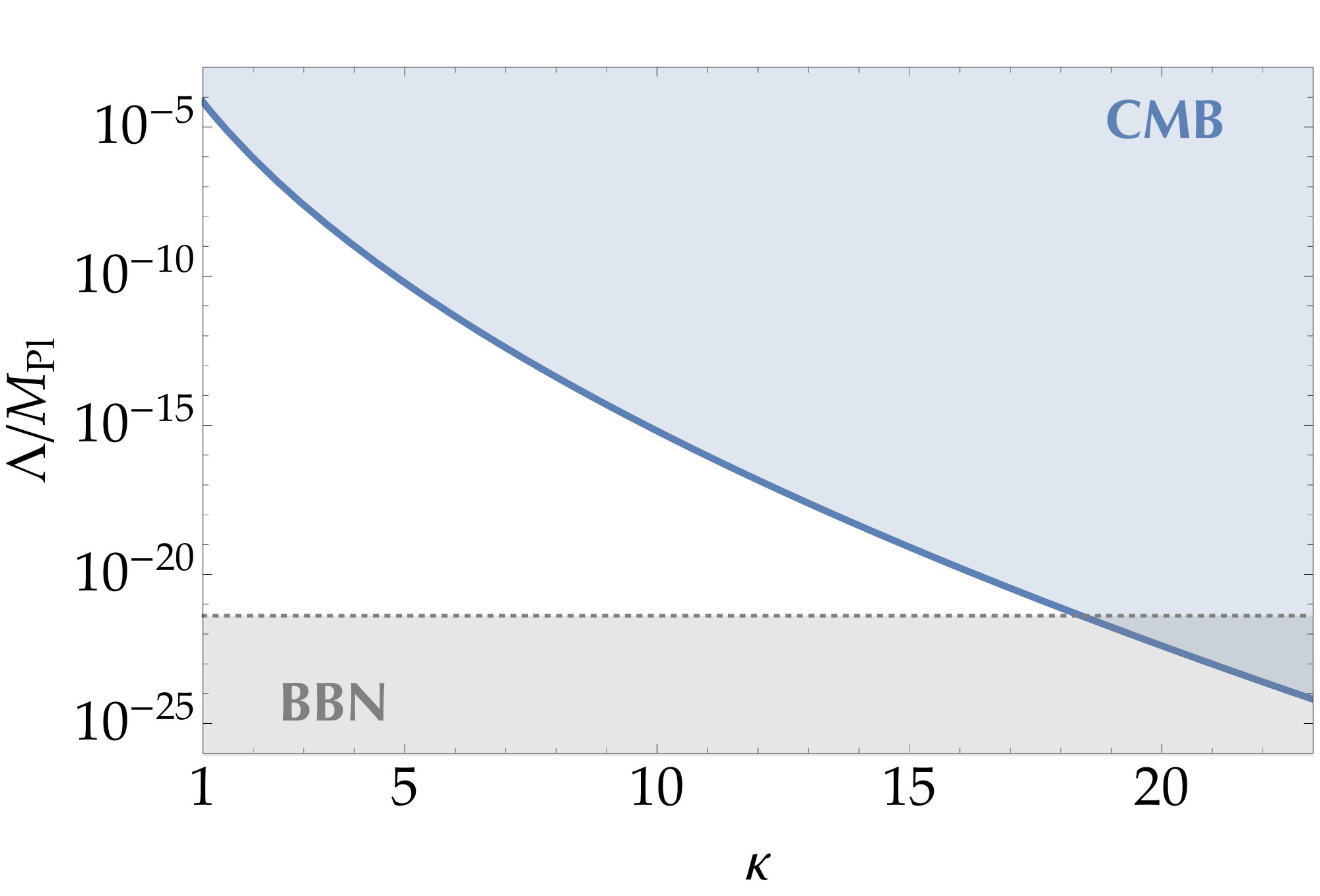}{The excluded region of the phase-0 (CMB scale) energy scale $\Lambda$ and the potential curvature $\kappa\simeq-\eta_V$. The blue line shows the upper bound on $\Lambda$ so that the phase-0 continues more than 15 e-folds and also the curvature perturbations generated by the inflaton $\phi_0$ are smaller than the observed value $\calP_{\zeta_{\phi_0}}\lesssim2\times10^{-9}$~\cite{Aghanim:2018eyx} to utilize the spectator scenario. The gray-dotted line is a lower limit by the BBN bound $\Lambda\gtrsim1\,\mathrm{MeV}$.}{fig: lambda kappa}

We finish this section by mentioning the TCC condition in multi-phase inflation.
If the universe follows the standard cosmology after the reheating, the horizon scale $H^{-1}$ grows faster than the comoving expansion $\propto a$ and therefore
the TCC condition~(\ref{eq: TCC}) does hold in the later universe once it holds at the reheating $t_R$:
\bae{
	\frac{a(t_R)}{a_\mathrm{ini}}\lpl<\frac{1}{H(t_R)}.
}
On the other hand, the current horizon scale $H_0^{-1}$ should be inside the horizon at the initial time of the first inflation as\footnote{One may consider the possibility that the current horizon scale was outside the horizon at the initial time, but entered the horizon during some long-lasting inflaton oscillation phase, and then reexited the horizon during the phase-0. However such a pre decelerated expansion strengthens the TCC constraint~\cite{Mizuno:2019bxy,Brandenberger:2019eni} and we do not consider such a scenario to relax the TCC condition.}
\bae{
	\frac{a_\mathrm{ini}}{a_0}H_0^{-1}<H^{-1}(t_\mathrm{ini}),
}
where $a_0$ represents the current scale factor.
Combining them, one obtains the constraint on the energy scale of the first inflation $H_\infl\simeq H(t_\mathrm{ini})$ as
\bae{
	\frac{H_\infl}{\Mpl}<\frac{12\sqrt{5}}{\sqrt{\pi}}\frac{g_{*s}^{1/3}(t_R)}{g_{*s}^{1/3}(t_0)g_*^{1/2}(t_R)}\frac{\Mpl H_0}{T_0T_R}\sim 300\frac{T_0}{T_R},
}
making use of the Friedmann equation $3\Mpl^2H^2=\frac{\pi^2}{30}g_*T^4$ and the entropy conservation $g_{*s}a^3T^3=\text{const.}$ with the radiation temperature $T$.
$g_*$ and $g_{*s}$ are the effective degrees of freedom for energy and entropy density.
For the last approximation, we use the current values $g_{*s}(t_0)\simeq3.93$, $T_0\simeq2.725\,\mathrm{K}$, and $H_0\simeq67\,\mathrm{km}\,\mathrm{s}^{-1}\,\mathrm{Mpc}^{-1}$~\cite{2009ApJ...707..916F,Aghanim:2018eyx}, and assume the standard model values $g_*(t_R)\sim g_{*s}(t_R)\sim106.75$ at the reheating.

If inflation is single-phase and the reheating is completed almost instantaneously as $\Lambda_\infl^4:=3\Mpl^2H_\infl^2\sim\frac{\pi^2}{30}g_*(t_R)T_R^4$,
the inflation energy scale is then severely constrained as~\cite{Bedroya:2019tba}
\bae{
	\frac{\Lambda_\infl}{\Mpl}\lesssim\left(72\sqrt{30}\frac{g_{*s}(t_R)^{2/3}}{g_{*s}(t_0)^{2/3}g_*(t_R)^{1/2}}\frac{H_0^2}{T_0^2}\right)^{1/6}\sim5\times10^{-10}.
}
However, if it is followed by multiple phases of inflation,\footnote{Of course, each phase should also satisfy the TCC condition~(\ref{eq: TCC}).}
the reheating temperature can be lowered to the BBN constraint $T_R\gtrsim1\,\mathrm{MeV}$.
Assuming the phase-0 is the first inflation without any preinflation (negative-$i$ phase), the constraint is thus much relaxed as~\cite{Mizuno:2019bxy}
\bae{
	\frac{\Lambda_\infl}{\Mpl}\lesssim3\times10^{-4}.
}
This is weaker than the constraint by the dS conjecture shown in Fig.~\ref{fig: lambda kappa}. Thus multi-phase inflation can be compatible also with TCC.

\section{Spectator in multi-phase inflation}\label{sec: curvaton scenario in multi-phase inflation}

We saw that multi-phase inflation can be consistent with several types of (not-to-be-in) swampland conditions simultaneously.
However either of the first or second slow-roll condition is always violated in this case and thus the primordial curvature perturbations generated by inflatons
inevitably have a significant scale-dependence inconsistently with the CMB observation because the spectral index is roughly evaluated by the summation of those slow-roll parameters:
\bae{\label{eq: inflaton ns}
	\ns-1=\dif{\log\calP_{\zeta_\phi}}{\log k}\approx-6\epsilon_V+2\eta_V.
}
In this section, we see that the spectator can instead have almost scale invariant perturbations.

The spectator $\sigma$ is a very light scalar field, compared to the Hubble scale during all the inflationary phases.
Though it does not affect the inflation dynamics, it also gets fluctuations $\delta\sigma\sim H/2\pi$ frozen for a while.
Well after inflation, its fluctuations can be converted to the adiabatic curvature perturbations in e.g. the curvaton or modulated reheating mechanism as we discuss in Sec.~\ref{sec: conclusions}.
Such a conversion can be parametrized as 
\bae{
    \calP_\zeta(k)&\simeq c^2\frac{\calP_{\delta\sigma}(k)}{\sigma^2},
}
where we define the conversion rate $c$ as $N_{\log\sigma}=\partial N/\partial(\log\sigma)$ in the context of the $\delta N$ formalism~\cite{Lyth:2004gb}. The combination of $\delta\sigma/\sigma$ is useful as it is almost time-independent after its horizon exit for the spectator field with the quadratic potential.
In addition, in the curvaton mechanism, $\delta\sigma/\sigma$ directly corresponds to the isocurvature perturbation $S=\zeta_\sigma-\zeta_{\rm r}$ \cite{Langlois:2004nn}.
In particular, this conversion rate $c$ in the curvaton mechanism is given by $2r/3$ with the energy fraction $r=3\rho_\sigma/(4\rho_r+3\rho_\sigma)$ of the spectator $\rho_\sigma$ to the background radiation $\rho_r$ at its decay time~\cite{Enqvist:2001zp,Lyth:2001nq,Moroi:2001ct}.
The modulated reheating scenario gives $c\simeq-\frac{1}{6}\pdif{\log\Gamma}{\log\sigma}$ with the (last) inflaton's decay rate $\Gamma$, the numerical factor $-1/6$ being varied by the inflaton's decay scenario~\cite{Ichikawa:2008ne}.
In these cases, the scale-dependence of the final curvature perturbation is determined only by the spectator perturbation.
If its (effective) mass $m_{\sigma,\eff}$ is not completely negligible during inflation, its perturbation is not fully frozen but leads to a scale-dependence in addition to the time evolution of $H$ as
\bae{
	\calP_{\delta\sigma}=\left(\frac{H_k}{2\pi}\right)^2\left(\frac{k}{a_\uf H_\uf}\right)^{2m_{\sigma,\eff}^2/3H^2},
}
where $H_k$ is the Hubble parameter at the time of the horizon exit $k=aH$ and the subscript $\uf$ indicates the end of (phase-0) inflation. The spectral index of $\calP_{\zeta_\sigma}$ is thus given by
\bae{
	\ns-1=\dif{\log\calP_{\zeta_\sigma}}{\log k}=-2\epsilon_H+\frac{2m_{\sigma,\eff}^2}{3H^2}.
}
Compared to the inflaton's case~(\ref{eq: inflaton ns}), it can be small enough even if $|\eta_V|>1$ as long as $\epsilon_H\sim\epsilon_V\ll1$.

In a multi-inflation scenario, we assume $\epsilon_V\ll1$ during each inflationary phase. Thus, in order to explain the observed value 
$\ns=0.965\pm0.004$~\cite{Aghanim:2018eyx}, the spectator mass is expected to be $m_{\sigma,\eff}^2\sim-0.05H^2$ during the phase-0.
However, in contrast to the ordinary case, the CMB scale inflation (phase-0) is followed by lower energy inflations.
Such a large tachyonic mass, in this case, lets the spectator roll down to and oscillate around its potential minimum, diluting its fluctuations.
The spectator $\sigma$ then cannot play the role of the perturbation source.
We instead assume that the tachyonic mass for $\sigma$ is dynamically yielded only during the phase-0.
The total potential of the system is given by
\bae{
	V=V_\infl+\frac{1}{2}m_\sigma^2\sigma^2-\frac{1}{2}c_{0\sigma}V_{\hill,0}\frac{\sigma^2}{\Mpl^2},
}
with a positive small coupling $c_{0\sigma}\sim0.02$ and the intrinsic mass $m_\sigma$ negligibly small during all phases of inflation.\footnote{Large absolute value of $c_{0\sigma}$ settles $\sigma$ down to the effective potential minimum 
and dilutes its perturbations even during the phase-0. Thus the perturbations given by the spectator scenario tend to be near scale-invariant.}
The spectator perturbation $\delta\sigma$ gets red-tilted due to the effective tachyonic mass 
$\partial_\sigma^2V\simeq-c_{0\sigma}V_{\hill,0}/\Mpl^2\simeq-3c_{0\sigma}H^2$ during the phase-0.
After the phase-0, the tachyonic mass decays together with $V_{\hill,0}$, keeping $\sigma$ from rolling down to the potential minimum.
In the curvaton scenario, $\sigma$ oscillates with its intrinsic mass $m_\sigma$ and increases its energy fraction to the background,
while the mass $m_\sigma$ is not necessary in the modulated reheating case.

\begin{figure*}[htbp]
    \centering
    \begin{tabular}{cc}
        \begin{minipage}{0.5\hsize}
            \includegraphics[width=0.9\hsize]{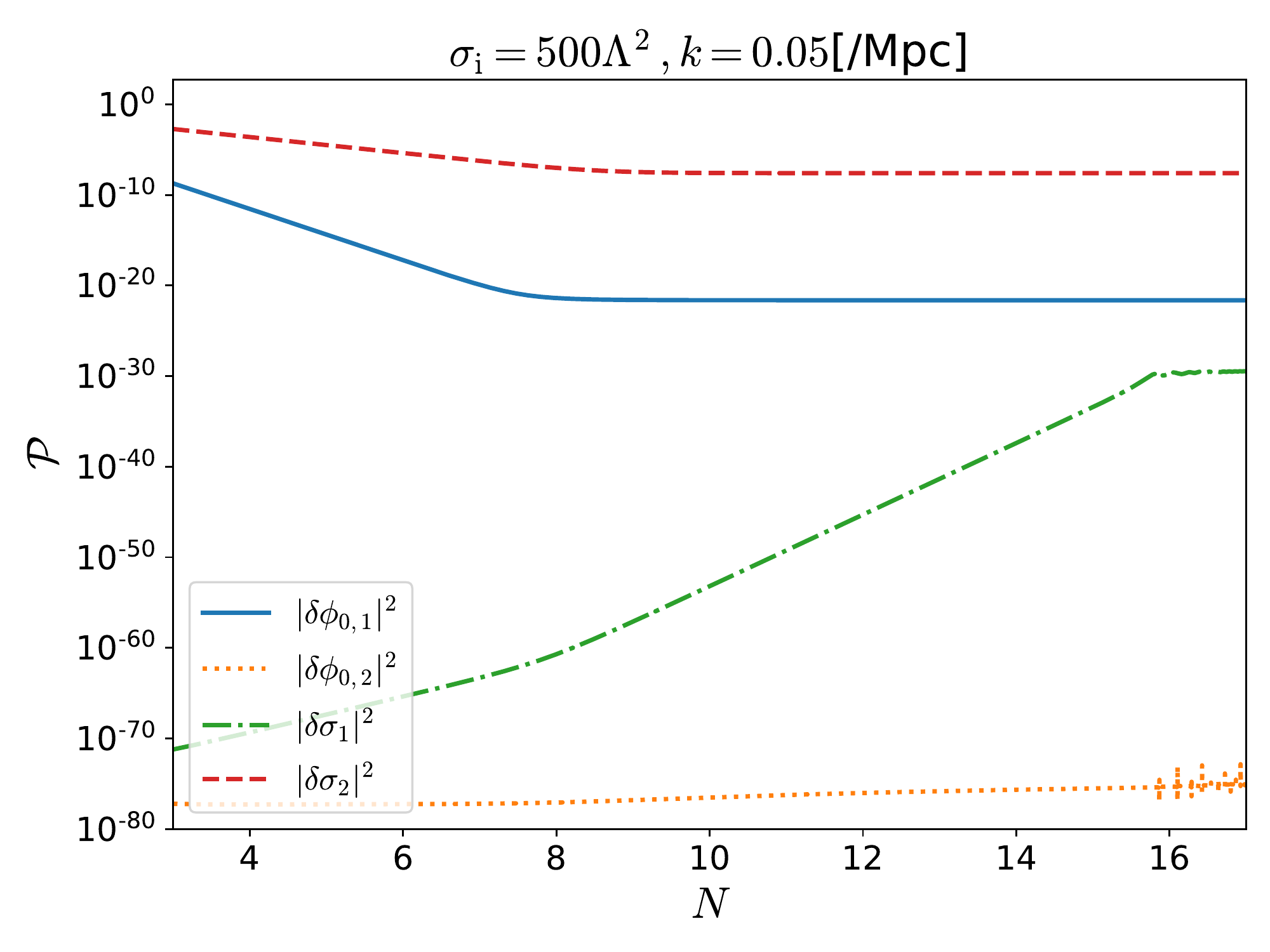}
        \end{minipage}
        \begin{minipage}{0.5\hsize}
            \includegraphics[width=0.9\hsize]{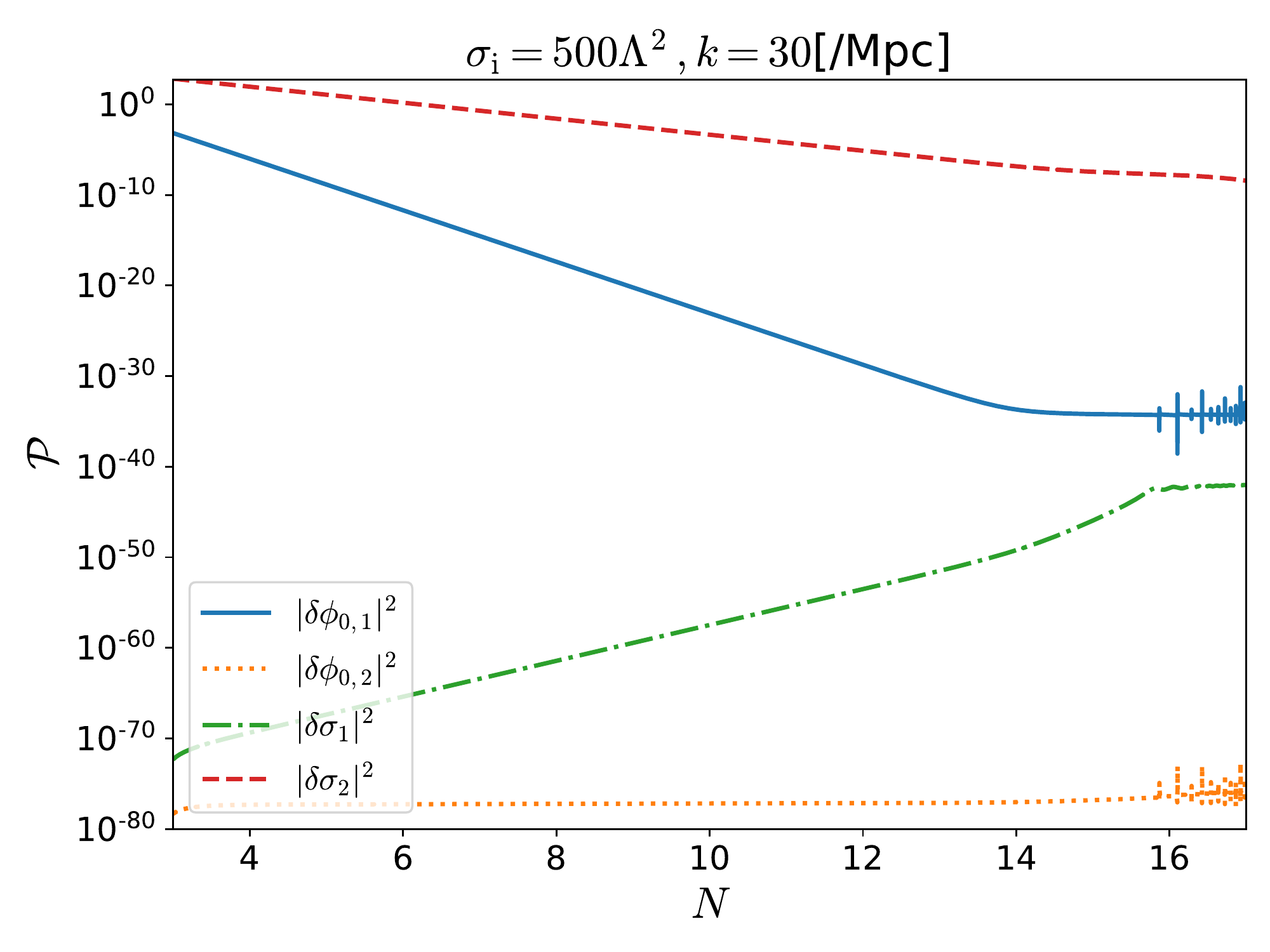}
        \end{minipage}
    \end{tabular}
    \caption{The time evolution of each perturbation~(\ref{eq: each calPzeta}) $\calP_{\zeta_{\phi_{0,1}}}$~(blue), $\calP_{\zeta_{\phi_{0,2}}}$~(orange dotted), $\calP_{\delta{\sigma_1}}/\sigma^2$~(green dot-dashed), and $\calP_{\delta{\sigma_2}}/\sigma^2$~(red dashed) for $k=0.05\,\mathrm{Mpc}^{-1}$ and $k=30\,\mathrm{Mpc}^{-1}$.
    The $k$'s normalization is fixed by hand to satisfy the observational constraints (see Fig.~\ref{fig: calPzeta}).
    Noisy features around $N\sim16$ simply originate from the numerical error and do not have any physical implication.
    Even the high frequency mode ($k=30\,\mathrm{Mpc}^{-1}$) around the horizon scale at the end of the phase-0 ($k_\uf\simeq45\,\mathrm{Mpc}^{-1}$) safely avoids a resonant amplification.}
    \label{fig: perturb}
\end{figure*}

Let us show some numerical results in a concrete model.
To see the dynamics during and after the phase-0, we specify the whole form of the phase-0 inflaton potential, instead of the expansion around the potential top, as
\bae{
	V_{\hill,0}=\Lambda^4\left(1-\frac{\phi_0^2}{v_0^2}\right)^2, 
}
respecting the distance conjecture~(\ref{eq: distance conjecture}) as $v_0\lesssim\Mpl$.\footnote{Though we choose $n=2$ here, the wine bottle potential can be generally described by $\propto(1-\phi_0^n/v_0^n)^2$ with an arbitrary power $n$. However such a potential often causes a resonant amplification in perturbations soon after inflation, easily losing the analytic predictability. According to the work in Ref.~\cite{Inomata:2017uaw}, $n\leq3$ and $v_0\gtrsim0.1\Mpl$ are favored to avoid the resonance. In our setup, any resonant feature is not shown either Fig.~\ref{fig: perturb} or Fig.~\ref{fig: calPzeta} and thus the resultant perturbations will not conflict with the observational constraints.}
Specifically we choose parameter values and initial conditions for $\phi_0$ and $\sigma$, $\phi_{0\ui}$ and $\sigma_\ui$, at the onset of the phase-0 as
\bae{\label{eq: parameters}
    \bce{
        \dps
	    \Lambda=10^{-9}\Mpl, \qquad v_0=\Mpl, \qquad c_{0\sigma}=0.02, \\
	    \dps
	    \phi_{0\ui}=360\Lambda^2/\Mpl, \qquad \sigma_{\ui}=500\Lambda^2/\Mpl,
	}
}
and the spectator's intrinsic mass $m_\sigma$ is neglected.
The background dynamics is shown in Fig.~\ref{fig: phi0 and sigma}. While the inflaton $\phi_0$ grows significantly due to its large tachyonic mass,
$\sigma$ is almost frozen even after the phase-0 because $\sigma$'s tachyonic mass decays together with the inflaton potential $V_{\hill,0}$.

\bfe{width=0.9\hsize}{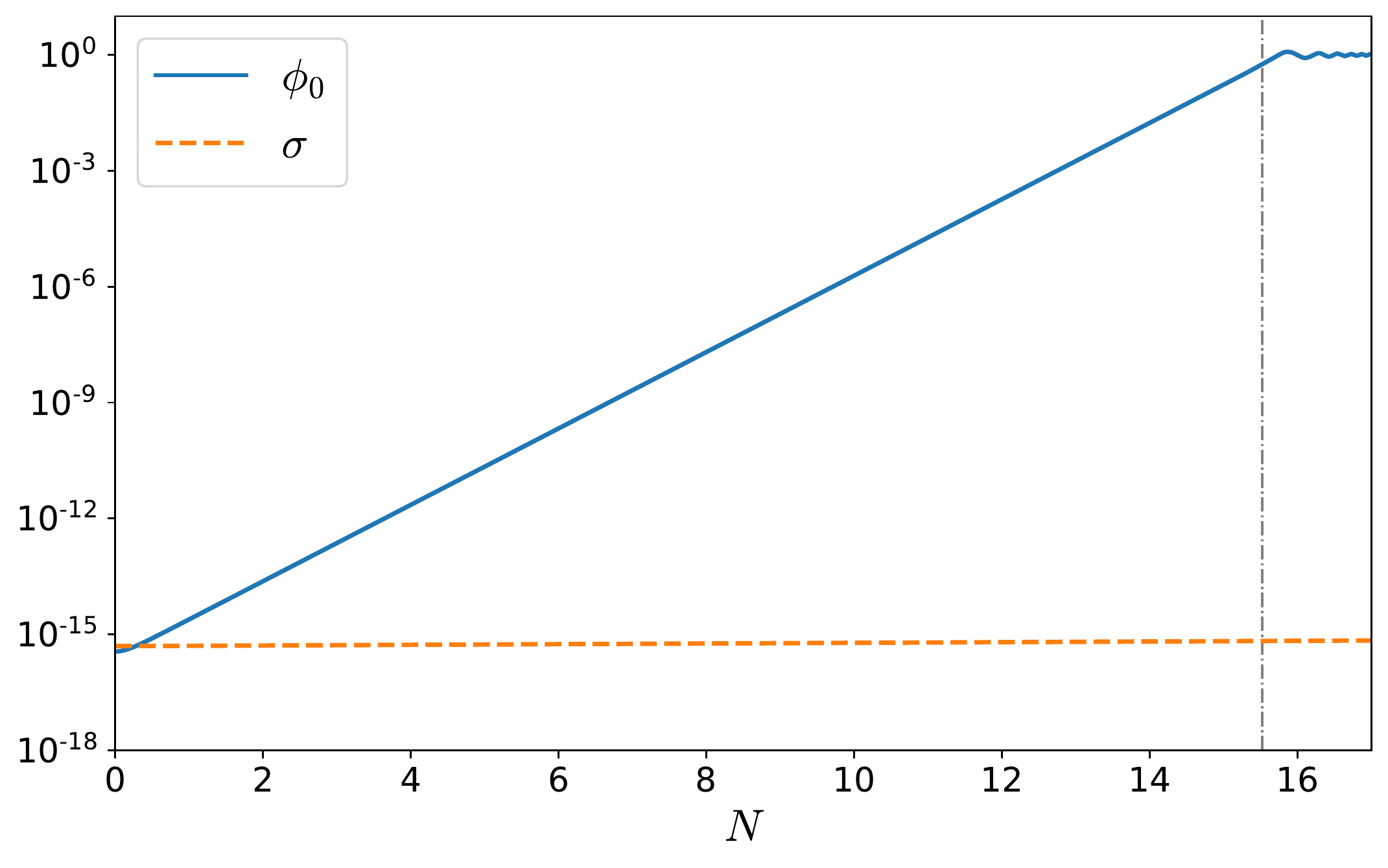}{The background dynamics of $\phi_0$ (blue) and $\sigma$ (orange dashed) in the unit of $\Mpl$. The vertical dot-dashed line represents the end of the phase-0.}{fig: phi0 and sigma}

Their perturbations can be obtained by solving linear Fourier-space EoM on the flat slice~\cite{Sasaki:1995aw}
\bae{
	&\delta\ddot{\phi}^I_\mathbf{k}+3H\delta\dot{\phi}^I_\mathbf{k}+\left(\frac{k^2}{a^2}\delta^I_J+\delta^{IK}V_{KJ}\right)\delta\phi^J_\mathbf{k} \nonumber \\
	&\qquad=\frac{1}{a^3\Mpl^2}\dif{}{t}\left(\frac{a^3}{H}\dot{\phi}^I\dot{\phi}^J\right)\delta_{JK}\delta\phi^K_\mathbf{k}.
}
Indices $I,J,K,\cdots$ label $\phi_0$ ($I=1$) or $\sigma$ ($I=2$). $V_{IJ}$ represents the potential second derivative $\partial_{\phi^I}\partial_{\phi^J}V$.
In the multi-field case, one has to consider the matrix mode function $\delta\phi^I_{\mathbf{k}\alpha}$ ($\alpha=1$ or 2)
because of the mode mixing through the non-diagonal parts of the Hessian $V^I{}_J=\delta^{IK}V_{KJ}$
and the gravitational interaction $\frac{1}{a^3\Mpl^2}\dif{}{t}\left(\frac{a^3}{H}\dot{\phi}^I\dot{\phi}^J\right)\delta_{JK}$.
Their initial condition can be chosen as
\bae{\label{eq: initial condition}
	\delta\phi^I_{\mathbf{k}\alpha}(t)\to\frac{\delta^I_\alpha}{a(t)\sqrt{2k}}\ee^{-ik\int a^{-1}\dd t}.
}

Together with the adiabatic perturbation by the inflaton $\zeta_{\phi_0}$, the spectator can make the final mixed curvature perturbation parametrized as
\bae{
    \zeta_\alpha=\zeta_{\phi_{0,\alpha}}+c\frac{\delta\sigma_{\alpha}}{\sigma},
}
where
\bae{
    \zeta_{\phi_{0,\alpha}}=-H\frac{\delta\phi_{0,\alpha}}{\dot{\phi}_0}.
}
Its power spectrum is then given by
\bae{\label{eq: mixed calPzeta}
    &\calP_\zeta=\sum_\alpha\left(\calP_{\zeta_{\phi_{0,\alpha}}}+c^2\frac{\calP_{\delta{\sigma_\alpha}}}{\sigma^2}+c\frac{\calP_{{\mathrm{mix},\alpha}}}{\sigma}\right), \\
    &\bce{
        \dps
        \calP_{\zeta_{\phi_{0,\alpha}}}=\frac{k^3}{2\pi^2}|\zeta_{\phi_{0,\alpha}}|^2, \\[5pt]
        \dps
        \calP_{\delta{\sigma_\alpha}}=\frac{k^3}{2\pi^2}|\delta{\sigma_\alpha}|^2, \\[5pt]
        \dps
        \calP_{{\mathrm{mix},\alpha}}=\frac{k^3}{2\pi^2}|\zeta_{\phi_{0,\alpha}}\delta{\sigma_\alpha}|.
    }\label{eq: each calPzeta}
}
The time evolution of each perturbation $\calP_{\zeta_{\phi_{0,\alpha}}}$ and $\calP_{\delta{\sigma_\alpha}}/\sigma^2$ is shown in Fig.~\ref{fig: perturb}.
The resultant power spectra are also exhibited in Fig.~\ref{fig: calPzeta}.
Here the conversion rate $c$ and the scale normalization are chosen by hand
so that the observational constraints are satisfied, assuming that the dynamics after the phase-0 is suitably realized (specifically $c=0.28$). 
Almost scale-invariant curvature perturbations over the enough range of scales are explained by the spectator scenario in multi-phase inflation.
This is the main result in this paper.

\bfe{width=1\linewidth}{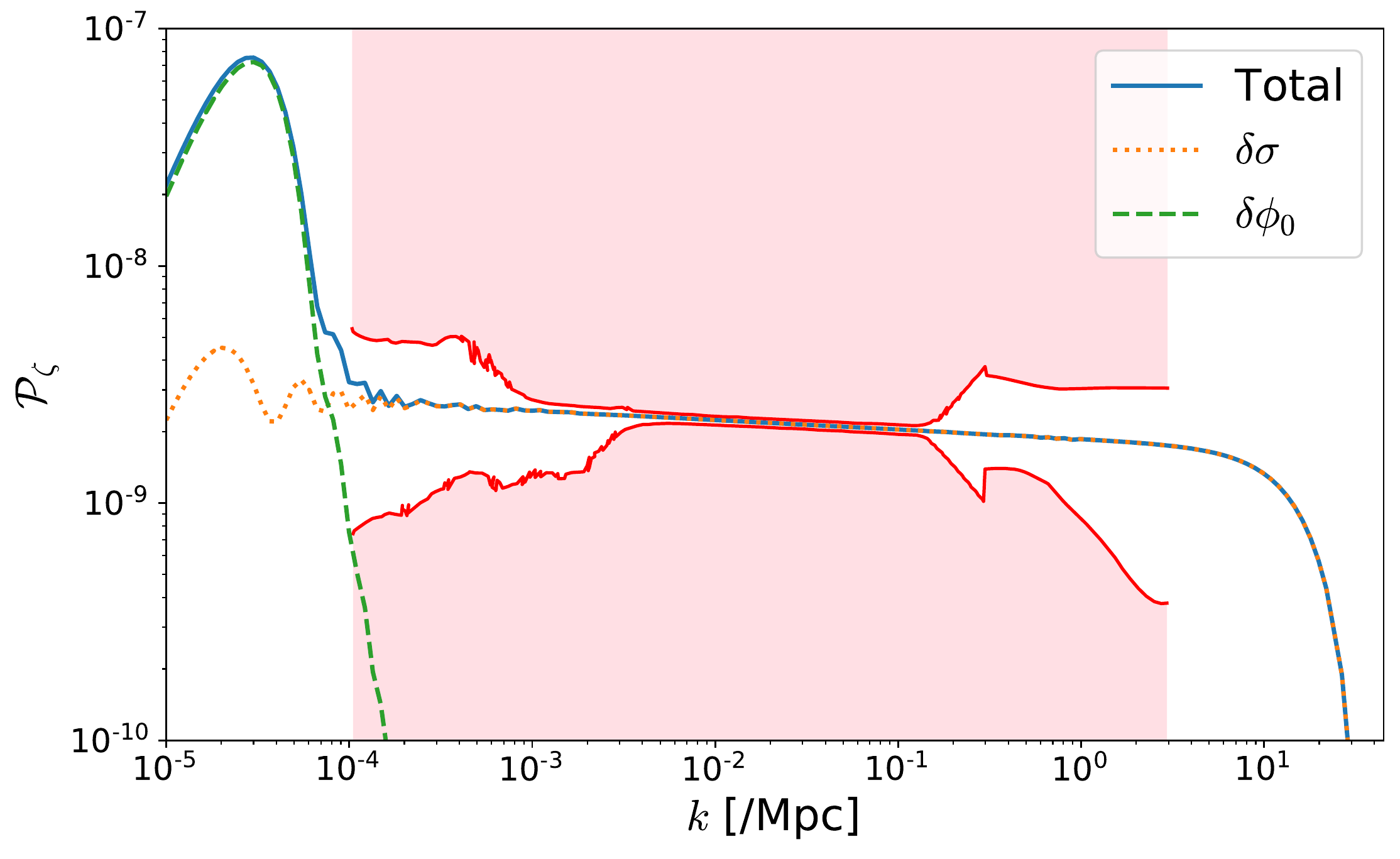}{The resultant power spectra of mixed curvature perturbations $\calP_\zeta$~(\ref{eq: mixed calPzeta}) (blue) as well as each mode $c^2\calP_{\delta\sigma}/\sigma^2$~(orange dotted) and $\calP_{\zeta_{\phi_0}}$~(green dashed). The conversion rate $c=0.28$ and the scale normalization is chosen by hand.
The red region is excluded by the CMB and LSS observations~\cite{Akrami:2018odb,Bird:2010mp}. The wiggling features of power spectra on the large scale $k\lesssim10^{-4}\,\mathrm{Mpc}^{-1}$ are simply reflecting the fact that these modes exit the horizon soon after the beginning of the phase-0 and cannot be well initialized by the deep subhorizon solution~(\ref{eq: initial condition}).
Their detailed form can be altered by the actual dynamics of preinflation (negative-$i$ phase).}{fig: calPzeta}

\section{Discussion and conclusion}\label{sec: conclusions}

In this paper, we point out that the multiple inflationary scenario can be compatible with the distance, dS, and trans-Planckian censorship conjectures with use of a spectator whose perturbations can be converted into the almost scale-invariant curvature perturbations on the CMB scale.
Let us first discuss the possible scenarios of such a perturbation conversion in this section.

The curvaton scenario~\cite{Enqvist:2001zp,Lyth:2001nq,Moroi:2001ct} is a famous mechanism to convert the spectator perturbation into the adiabatic mode.
Even if the spectator's energy fraction is quite tiny at first, once it starts to oscillate with its mass term, it behaves as a matter fluid and its relative energy density to the background radiation can grow as time goes.
When the spectator decays into radiations, its perturbations are converted to the adiabatic curvature perturbations with the conversion rate given by the energy fraction $r=3\rho_\sigma/(4\rho_r+3\rho_\sigma)$ at that time.
The interesting feature of the curvaton mechanism is that the conversion rate $r$ is directly related with the non-Gaussianity of the resultant curvature perturbations.
In terms of the non-linearity parameter $\fNL$, the relation is given by~\cite{Lyth:2005du}
\bae{
    \fNL\simeq-\frac{5}{3}-\frac{5}{6}r+\frac{5}{4r},
}
neglecting the inflaton's contribution.
As the CMB observation by the Planck collaboration constrained this non-linearity parameter as $\fNL=-0.9\pm5.1$~\cite{Akrami:2019izv},
the curvaton should have a non-negligible energy fraction at its decay time as $0.21\lesssim r\leq1$.

However the swampland conditions make it harder for the curvaton to dominate the universe.
It is caused by the low energy scale $\Lambda$ of inflation, which determines the amplitude of the curvaton fluctuations by $\delta\sigma\sim H/2\pi\sim\Lambda^2/2\sqrt{3}\pi\Mpl$.
As the curvaton is assumed to be the source of the CMB scale adiabatic perturbation $\zeta\sim5\times10^{-5}$~\cite{Aghanim:2018eyx}, it also fixes the relation between the background field value $\sigma$ and the inflation energy scale $\Lambda$ as $\sigma\sim10^3\Lambda^2/\Mpl$, as can be seen in our parameters~(\ref{eq: parameters}).
On the other hand, at the onset of the curvaton oscillation $H\sim m_\sigma$, its energy fraction to the background radiation can be expressed as
\bae{
	\left.\frac{\rho_\sigma}{\rho_r}\right|_\mathrm{osc}\sim\left.\frac{m_\sigma^2\sigma^2}{H^2\Mpl^2}\right|_\mathrm{osc}
	\sim\frac{\sigma^2}{\Mpl^2}\sim10^6\left(\frac{\Lambda}{\Mpl}\right)^4,
}
which is extremely suppressed in low-scale inflation. For example, our choice of parameters~(\ref{eq: parameters}) reads $\rho_\sigma/\rho_r|_\mathrm{osc}\sim2.5\times10^{-31}$.
It only grows as the scale factor $a\propto T^{-1}$, obviously indicating that the curvaton cannot dominate the universe well before the BBN era $T\sim 1\,\mathrm{MeV}$.
Therefore the curvaton paradigm is in tension with low-scale (landscape) inflation.
One may flatten the curvaton potential to delay the onset of the curvaton oscillation as $H_\mathrm{osc}\ll m_\sigma$.
In this case, however the non-Gaussianity tends to be large because the oscillation onset itself depends on the fluctuation~\cite{Kawasaki:2011pd}.

One can also convert perturbations by varying the inflaton's decay through the spectator field, known as the modulated reheating scenario~\cite{Dvali:2003em,Kofman:2003nx}.
For example, if the (last) inflaton $\phi_\ell$ decays into the descendent fermions $\psi$ through the Yukawa interaction $y\phi_\ell\bar{\psi}\psi$, it can be corrected by higher dimension couplings as $\alpha^\prime\frac{\sigma}{M}\phi_\ell\bar{\psi}\psi+\beta^\prime\frac{\sigma^2}{M^2}\phi_\ell\bar{\psi}\psi+\cdots$ with some cutoff scale $M$.
The modulated decay rate is then parametrized as
\bae{
    \Gamma=\Gamma_0\left(1+\alpha\frac{\sigma}{M}+\beta\frac{\sigma^2}{M^2}+\cdots\right),
}
$\alpha$ and $\beta$ would be order unity coefficients and the cutoff scale is assumed to be larger enough than the spectator's background value as $M\gg\sigma$.
If the inflaton $\phi_\ell$ oscillates by the quadratic potential before its decay, the conversion rate is given by $c=-\frac{1}{6}\pdif{\log\Gamma}{\log\sigma}$ in this case~\cite{Ichikawa:2008ne}.
At the leading order, the resultant adiabatic perturbation can be evaluated as 
$\zeta\sim\alpha\frac{\delta\sigma}{M}\sim\frac{\Lambda^2}{2\sqrt{3}\pi M\Mpl}$, which should be $\sim5\times10^{-5}$.
Thus the cutoff scale will be $M\sim10^3\Lambda^2/\Mpl$.
This is relatively small ($\sim1\,\mathrm{TeV}$ in our setup~(\ref{eq: parameters})) but may be possible.
The background spectator value $\sigma$ should be smaller than our choice to satisfy $M\gg\sigma$ in this case.

The non-Gaussianity in the modulated reheating scenario is also controllable.
If the inflaton oscillates by the quadratic potential and decays through the Yukawa interaction, the non-linearity parameter reads~\cite{Ichikawa:2008ne}
\bae{\label{eq:modulated fnl}
    \fNL\simeq5\left(1-\frac{\Gamma\Gamma_{\sigma\sigma}}{\Gamma_\sigma^2}\right)
    \simeq 5 \left( 1 -\frac{\beta}{\alpha^2} \right),
}
neglecting the inflaton's contribution.\footnote{Eq.~(\ref{eq:modulated fnl}) can be applied only if the spectator's background dynamics is negligible after inflation like as our case. Otherwise the non-linearity parameter can be changed~\cite{Kobayashi:2013nwa}.}
Here $\Gamma_\sigma=\partial\Gamma/\partial\sigma$ and $\Gamma_{\sigma\sigma}=\partial^2\Gamma/\partial\sigma^2$.
Such an order unity non-Gaussianity can be compatible with the current constraint $\fNL=-0.9\pm5.1$~\cite{Akrami:2019izv}, and moreover can be detectable with future galaxy surveys as SPHEREx~\cite{Dore:2014cca,Dore:2018kgp}, LSST~\cite{Abell:2009aa}, and Euclid~\cite{Laureijs:2011gra} and/or 21cm observations like SKA~\cite{Maartens:2015mra} for example.
We leave further discussions about the conversion mechanism and the resultant non-Gaussianity for future works.

Let us also mention the smaller scale perturbations as another interesting feature of our scenario other than the non-vanishing non-Gaussianity.
They can be completely different from those on the CMB scale as they correspond with different phases of inflation.
If the same spectator is responsible also for these small scale perturbations, their amplitudes will decrease stepwise because the spectator's perturbations are proportional to the energy scale of each phase of inflation.
Currently the small scale primordial perturbations have been constrained only with the upper bound by the non-detection of primordial black holes (PBHs)~\cite{Carr:2020gox} or ultracompact minihalos~\cite{Bringmann:2011ut}.
However too little perturbations on $\sim10^{-3}\text{--}10^{-1}\,\mathrm{Mpc}$ may delay the early structure formation and thus change the reionization history, which can be probed by future 21cm observations~\cite{Yoshiura:2019zxq}. In such a way, one may impose a lower limit on the small scale perturbation as another consistency check of our scenario.

On a smaller scale, inflaton also can make a dominant contribution to the curvature perturbation. As can be seen in Fig.~\ref{fig: calPzeta}, the curvature perturbation has a significant scale-dependence in this case due to the violation of the slow-roll condition.
In other words, the power spectrum of the curvature perturbation can have a peak on some scale. If the perturbation amplitude is large enough at such a peak, PBHs can be formed and may explain the dark matter or gravitational waves detected by the LIGO/Virgo collaboration as suggested in Ref.~\cite{Tada:2019amh}. We also leave these possible detectabilities for future works.

\acknowledgments

We are grateful to Fuminobu Takahashi and Shuichiro Yokoyama for helpful advice to the main idea of this work. We also thank Takeshi Kobayashi and Ryo Saito for useful discussions.
This work is supported by JSPS KAKENHI Grants No. 
JP18J01992 (Y.T.), No. JP19K14707 (Y.T.),
and No. JP19J22018 (K.K.)

\bibliography{main}
\end{document}